\documentclass[acus]{JAC2000}

\usepackage{graphicx}

\begin{document}
\title{{\uppercase{\boldmath
A HIGHLY FLEXIBLE BUNCH COMPRESSOR\\FOR THE APS LEUTL FEL}}
\thanks{
Work supported by the U.S. Department of Energy, Office of Basic Energy
Sciences, under Contract No. W-31-109-ENG-38.}}

\author{M. Borland, J. Lewellen, S. Milton, ANL, Argonne, IL 60439, USA}

\maketitle

\begin{abstract} 

The Low-Energy Undulator Test Line (LEUTL) free-electron laser
(FEL) \cite{MiltonFEL} at the Advanced Photon Source (APS) has achieved
gain at 530 nm with an electron beam current of about 100
A \cite{MiltonSASE,SerenoLINAC2K}.  In order to push to 120 nm and
beyond, we have designed and are commissioning a bunch compressor
using a four-dipole chicane at 100-210 MeV to increase the current to
600 A or more.  To provide options for control of emittance growth due
to coherent synchrotron radiation (CSR), the chicane has variable $R_{56}$.
The symmetry of the chicane is also variable via longitudinal motion
of the final dipole, which is predicted to have an effect on emittance
growth \cite{BorlandLINAC2K}.  Following the chicane, a three-screen
emittance measurement system should permit resolution of the
difference in emittance growth between various chicane configurations.
A vertical bending magnet analysis line will permit imaging
of correlations between transverse and energy
coordinates \cite{Dowell}.    A companion paper discusses the
physics design in detail \cite{BorlandLINAC2K}.
\end{abstract}

\section{APS LINAC OVERVIEW}

\begin{figure*}[htb]
\centering
\includegraphics*[width=140mm]{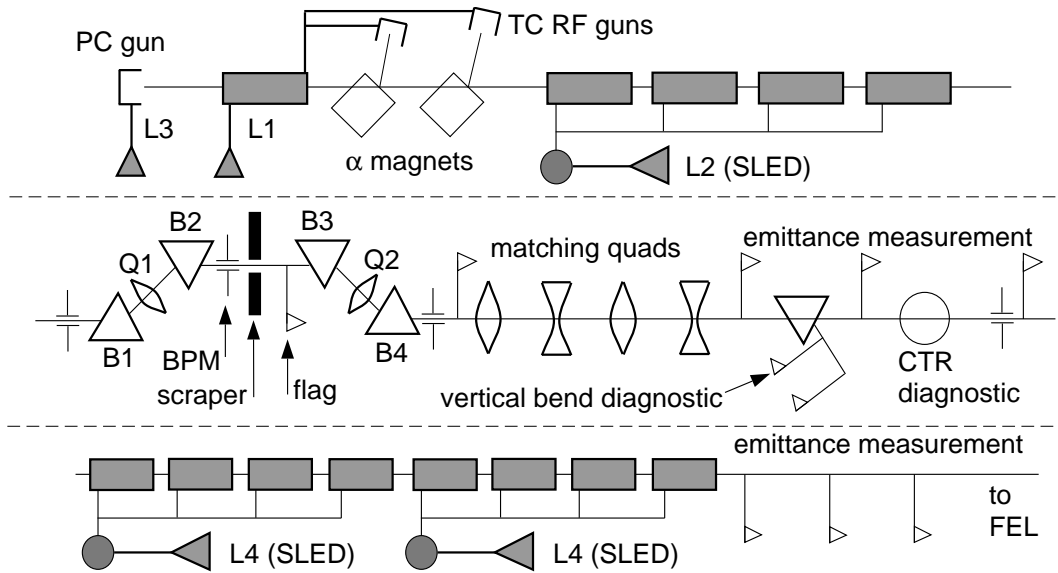}
\caption{Schematic of the APS linac with the bunch compressor.}
\label{fig:layout}
\end{figure*}

The APS injector consists of a linac, an accumulator ring, and a 7-GeV
booster synchrotron.  In addition to delivering beam to the
accumulator, the linac can be configured \cite{Soliday} to deliver
beam to the LEUTL experiment hall \cite{MiltonFEL}.  The linac
consists of 13 Stanford Linear Accelerator Center (SLAC) type accelerating
sections powered by four klystrons, two thermionic rf guns (TRFG)
\cite{BorlandThesis,BorlandRFG,LewellenRFG} powered (one at a time) by
a single klystron, and one photocathode gun (PCG) \cite{BiedronPCG}
powered by a single klystron.  Figure \ref{fig:layout} shows a
schematic of the system and the location of the newly-installed bunch
compressor.

The original purpose of the linac was to create positron beams and
deliver them to the accumulator ring for injection into the APS.  The
positron target was subsequently removed when the APS switched to
electron operation.  In both situations, the requirements on the linac
were modest in terms of emittance, energy spread, bunch length, and
stability.  However, the requirements for reliability were and are
very high, which was one reason for elimination of positron operation.
The FEL project requires much higher beam quality and beam stability.
The required beam quality is typically only achieved using a
photocathode gun; however, the reliability of such guns (particularly
the drive laser) is insufficient to act as an injector for the APS.
The dual thermionic guns have a distinct advantage here, having proven
themselves as components of the injector at SSRL \cite{WeaverRFG}.  The
use of alpha magnets \cite{BorlandThesis} for magnetic bunch compression in
these guns allows the guns to be placed off-axis, leaving the on-axis
position for the PCG.  This is an important consideration in
preserving the PCG beam brightness.

\section{MAGNETIC BUNCH COMPRESSION}

The principle of magnetic bunch compression is well-known, so we only
review the basic idea here.  In a magnetic chicane (see Figure
\ref{fig:layout}) the path length traveled by a particle is $s = s_o +
R_{56}\delta$, where $s_o$ is the central path length and $\delta =
(p-p_o)/p_o$ is the fractional momentum deviation.  For simple
chicanes, $R_{56}<0$ so that high-energy particles take a shorter path.

Phasing the beam ahead of the crest in the precompressor linac
introduces an ``energy chirp'' into the beam, so that the tail of the
beam has higher energy than the head.  As a result, the tail will
catch up to the head in the chicane, giving a shorter bunch.  If the
beam is undercompressed, then the energy spread imparted in the
precompressor linac can be removed by phasing behind the crest in the
postcompressor linac.

It is possible to derive formulae for the phasing required to obtain a
desired bunch length and minimized energy spread.  However, accurate
calculation requires including wakefield effects and depends on the
detailed initial bunch shape.  Hence, we used simulation to find the
optimal values \cite{BorlandLINAC2K}.

\section{LEUTL BEAM REQUIREMENTS}

The primary goal of the bunch compressor is to provide higher current beam to
the LEUTL FEL.  A secondary goal is characterization of CSR effects.
The bunch compressor was designed with two LEUTL operating points in
mind.  These operating points, distinguished primarily by the beam
current of 300 or 600 A, are summarized in Table \ref{tab:LEUTLreq}.
The requirements for charge and emittance are not difficult compared
to the state-of-the-art for photoinjector systems.  We hope that these
parameters can be achieved repeatably and easily to provide for
routine and stable operation.

Because of the very non-Gaussian longitudinal phase-space
distributions one typically sees in the compressor, we use the
following definition for the beam current:
\begin{math}
I_{80} = \frac{0.8 * Q_{total}}{\Delta t_{80}}
\end{math}
where $Q_{total}$ is the total charge in the beam
and $\Delta t_{80}$ is the length in time of the central
80\% of the beam.  The value of 80\% was used because this
includes most of the particles but typically excludes high-current
spikes that tend to occur at the head and tail.  Also, when we
refer to bunch length, we mean $\Delta t_{80}$.

Because the initial emittance is relatively large, it is desirable
that compression not make it larger.  For the 600-A case, however,
simulations \cite{BorlandLINAC2K} predict an emittance growth of up to
40\% due to CSR.  Hence, this part of the LEUTL requirement may not be
met.

\begin{table}[h] 
    \caption{Desired LEUTL Operating Points}
    \label{tab:LEUTLreq}
    \begin{center}
        \begin{tabular}{|l|c|c|c|}\hline
                              & Nominal & 300 A & 600 A \\ \hline
        Current (A)           &  100    & 300  & 600 \\ \hline
        Energy (MeV)          &  217    & 217  & 457 \\ \hline
        RMS en. spread (\%)  &  \verb|<|0.1   & \verb|<|0.1 & \verb|<|0.15 \\ \hline
        Initial charge (nC) &  0.5    & 0.5  & 0.5 \\ \hline
        Final charge (nC)   & 0.5 & 0.42 & 0.42 \\ \hline
        $\Delta t_{80}$ (ps) & 4  & 1.1 & 0.55 \\ \hline
        Norm. emittance ($\mu$m) & 5 & 5 & 5 \\ \hline
        Light wavelength (nm) & 530 & 530 & 120 \\ \hline
        \end{tabular}
    \end{center}                                    
\end{table}

\section{BUNCH COMPRESSOR FEATURES}

Figure \ref{fig:layout} provides a detailed schematic of the
compressor chicane.  One sees that most of the beam energy at the
entrance to the bunch compressor is due to the ``L2'' sector of the
linac, which consists of a single SLEDed klystron driving four
SLAC-type 3-m structures, delivering a beam energy of up to 210 MeV.
The bunch compressor was designed with the range from 100-210 MeV in
mind.

Table \ref{tab:BCParam} shows some of the principle parameters of the
bunch compressor.  A noteworthy feature of the APS design is that the
$R_{56}$ is designed to be variable, which will be accommodated
through transverse motion of the central dipole pair (B2 and B3 in
Figure \ref{fig:layout}).  This permits variation of the bending angle
without having to design magnets with large good field regions.  As a
result, we can vary $R_{56}$ between 0 and -65 mm.  Presently, due to
delivery problems with the flexible chambers, the chicane is installed
with fixed chambers.  Later this year we will install flexible curved
chambers in all the dipoles and telescoping chambers between the
dipoles.  The hardware required for motion of the magnets 
is already in place.

\begin{table}[h]
        \caption{Bunch Compressor Parameters} 
        \label{tab:BCParam}
        \begin{center}
        \begin{tabular}{|l|c|}\hline
        Maximum bend angle & 13.5$^\circ$ \\ \hline
        Maximum bend field & 0.86 T \\ \hline
        Effective bend length & 192 mm \\ \hline
        Maximum $R_{56}$         & -65 mm \\ \hline
        Maximum transverse motion & 184 mm \\ \hline
        Maximum longitudinal motion & 602 mm \\ \hline
        \end{tabular}
        \end{center}
\end{table}        

The symmetry of the chicane will also be variable through longitudinal
motion of the final dipole, B4.  The ratio of the B3-B4 distance to
the B1-B2 distance will be variable from 1.0 to 2.0, corresponding to
variations in the ratio of the angle of B1 to the angle of B4 from 1 to
1.8.  Two ``tweaker'' quads are required within the chicane to allow
matching the dispersion for asymmetric configurations.

Variable $R_{56}$ and symmetry is thought to be interesting in that
the effect of CSR should change with these parameters (or, more
fundamentally, with the bending angles).  The asymmetric
configurations have weaker bending in B3 and B4, where the beam is
shortest, which should decrease CSR effects.  However, these
configurations also have a larger drift between B3 and B4, which
allows CSR more room to act.  Simulations show a slight benefit to the
asymmetric configuration for 600 A, and greater benefit beyond that.
We hope to test these predictions once the flexible chambers and
emittance diagnostics are fully implemented.

At present, no attempt has been made to shield CSR by placing
small-gap chambers in the dipoles.  Our intention is to add this
feature if we find it necessary and to thus measure the effect.

\section{DIAGNOSTICS}

Because of concerns about CSR and jitter effects in the bunch compressor,
we have planned for extensive diagnostics for the system.  Although not
all diagnostics are completed at this time, we expect completion this
year.  Figure \ref{fig:layout} shows most of the planned diagnostics.

There are BPMs upstream and downstream of the chicane, plus one in the
center of the chicane that will give information on the energy
centroid.  This new design is monopulse-receiver-based and should have
single-shot resolution and reproducibility of 15 $\mu$m for charge of 0.1
to 2 nC.

The compressor will have a total of seven beam-imaging flags.  One
flag is in the chicane center, downstream of the two-blade beam
scraper.  Another is at the exit of B4, where a small horizontal
beamsize is required to minimize CSR effects.  Three flags with 1-m
spacing provide a three-screen emittance measurement system.  Several
of these flags use a new design incorporating two cameras---one for
low magnification and another for high magnification.  The high
magnification cameras should achieve a beam size resolution of 7 to 15
$\mu$m, depending on charge sensitivities.

The chicane bends the beam in the horizontal plane.  A vertical
spectrometer magnet is installed downstream of the chicane with two
flags.  The first flag allows imaging the $x-\delta$ correlations in
the beam \cite{Dowell}, which should give information on the effects of
CSR and wakes.  The second flag is used for energy spread and centroid
resolution.

For bunch length measurements, we will initially use a coherent
transition radiation (CTR) diagnostic \cite{SerenoLINAC2K}.  This
diagnostic has been successfully used with one of the TRFGs and showed
features on the 100-fs scale.  We have also left space for synchrotron
light ports on all of the dipoles and may use this radiation for
bunch-length measurements in the frequency domain \cite{YangPC}.

\section{FUTURE DEVELOPMENTS}

We are also interested in use of the bunch compressor with the TRFGs.
Bunch lengths of 350 fs have been obtained with one of these guns,
using alpha-magnet-based compression \cite{SerenoLINAC2K}.
Simulations predict that by also using the bunch compressor, bunch
lengths of 5-10 fs are possible with currents on the order of 500 A.
While this is not useful for FEL work, it may be useful to those
interested in ultrashort pulses.

We are also planning an energy upgrade to the APS linac to allow
energy of up to 1 GeV.  The present limit with the bunch compressor is
about 600 MeV.

\section{ACKNOWLEDGEMENTS}

We would like to acknowledge valuable suggestions from and calculations
done by Paul Emma and Vinod Bharadwaj, both of SLAC.  Their technical
note \cite{Emma} provided a valuable starting point for our design.

\end{document}